\documentclass[a4paper,fleqn,usenatbib]{mnras}

\usepackage{newtxtext,newtxmath}

\usepackage[T1]{fontenc}
\usepackage{ae,aecompl}

\usepackage{graphicx}	
\usepackage{amsmath}	
\usepackage{amssymb}	
\usepackage{float}
\usepackage{subfig}
\usepackage{rotating}

\usepackage{etoolbox}
\makeatletter
\patchcmd\@combinedblfloats{\box\@outputbox}{\unvbox\@outputbox}{}{%
  \errmessage{\noexpand\@combinedblfloats could not be patched}%
}%
\makeatother

\newcommand{\xmm}{\emph{XMM-Newton}}
\newcommand{\cha}{\emph{Chandra}}
\newcommand{\ros}{\emph{ROSAT}}

\title[Neutron star mass/radius constraints]{The radius of the quiescent neutron star in the globular cluster M13}

\author[A. W. Shaw et al.]{
A. W. Shaw$^{1}$\thanks{E-mail: aarran@ualberta.ca (AWS)}
C. O. Heinke,$^{1}$
A. W. Steiner,$^{2,3}$
S. Campana,$^{4}$
H. N. Cohn,$^{5}$
\newauthor W. C. G. Ho,$^{6,7}$
P. M. Lugger,$^{5}$
and M. Servillat$^{8}$
\\
$^{1}$Department of Physics, University of Alberta, CCIS 4-181, Edmonton, AB T6G 2E1, Canada\\
$^{2}$Department of Physics and Astronomy, University of Tennessee, Knoxville, TN 37996, USA\\
$^{3}$Physics Division, Oak Ridge National Laboratory, Oak Ridge, TN 37831, USA\\
$^{4}$INAF - Osservatorio astronomico di Brera, Via E. Bianchi 46, I-23807 Merate (LC), Italy\\
$^{5}$Department of Astronomy, Indiana University, 727 E. Third Street, Bloomington, IN 47405, USA\\
$^{6}$Department of Physics and Astronomy, Haverford College, 370 Lancaster Ave, Haverford, PA 19041, USA\\
$^{7}$Mathematical Sciences, Physics \& Astronomy and STAG Research Centre, University of Southampton, Southampton SO17 1BJ, UK\\
$^{8}$LUTH, Observatoire de Paris, PSL Research University, CNRS, Université Paris Diderot, 92190 Meudon, France
}

\date{Accepted XXX. Received YYY; in original form ZZZ}

\pubyear{2018}

\begin{document}
\label{firstpage}
\pagerange{\pageref{firstpage}--\pageref{lastpage}}
\maketitle

\begin{abstract}
X-ray spectra of quiescent low-mass X-ray binaries containing neutron stars can be fit with atmosphere models to constrain the mass and the radius. Mass-radius constraints can be used to place limits on the equation of state of dense matter. We perform fits to the X-ray spectrum of a quiescent neutron star in the globular cluster M13, utilizing data from \ros, \cha\ and \xmm, and constrain the mass-radius relation. Assuming an atmosphere composed of hydrogen and a 1.4${\rm M}_{\odot}$ neutron star, we find the radius to be $R_{\rm NS}=12.2^{+1.5}_{-1.1}$ km, a significant improvement in precision over previous measurements. Incorporating an uncertainty on the distance to M13 relaxes the radius constraints slightly and we find $R_{\rm NS}=12.3^{+1.9}_{-1.7}$ km (for a 1.4${\rm M}_{\odot}$ neutron star with a hydrogen atmosphere), which is still an improvement in precision over previous measurements, some of which do not consider distance uncertainty. We also discuss how the composition of the atmosphere affects the derived radius, finding that a helium atmosphere implies a significantly larger radius. 
\end{abstract}




\section{Introduction}

One of the unanswered questions in astrophysics concerns the interior physics of neutron stars (NSs), primarily the equation of state (EOS) of dense matter. The EOS, which also defines the relationship between mass and radius, is universal across all NSs \citep{Lattimer-2001}. However, constraining this relation is difficult due to the complications involved with measuring NS radii. It is possible to derive constraints on the radius of a NS ($R_{\rm NS}$) from spectral fits to their thermal X-ray emission, for example through spectroscopic observations of thermonuclear bursts \citep[e.g.][]{Ozel-2009,Ozel-2012a,Poutanen-2014,Ozel-2016a,Nattila-2017}. Pulse profile modelling of rotation-powered pulsars can also provide an independent, non-spectroscopic method of constraining $R_{\rm NS}$ \citep{Watts-2016,Ozel-2016b}. One key method of deriving EOS constraints has been through the fitting of X-ray spectra of quiescent low mass X-ray binaries (qLMXBs) containing NSs.


These systems typically exhibit soft X-ray spectra consisting of a thermal, blackbody-like component, sometimes with a harder, non-thermal component \citep{Campana-1998}. The nature of the thermal component has been debated over the last few decades. \citet{Brown-1998} claimed that the soft X-ray component can be explained by the `leakage' of heat deposited in the core during accretion episodes. This `deep crustal heating' model can predict the thermal spectrum of many qLMXBs \citep[e.g.][]{Rutledge-2001a}, though not for those which exhibit short accretion episodes, in which case the heat must be released at a shallower depth \citep[][]{Degenaar-2011b}. Some qLMXBs may continue to accrete at a low level, somewhat mimicking the spectrum of deep crustal heating \citep{Zampieri-1995}.

Regardless of the mechanism powering the soft component, the X-rays originate from the atmosphere of the NS. After accretion ceases, the elements stratify very quickly, leaving the lightest at the top \citep{Alcock-1980,Romani-1987}. Thus, studies of qLMXBs have often found that the soft X-ray spectra can be well described by a hydrogen atmosphere \citep[e.g.][]{Rutledge-2001a,Rutledge-2001b,Heinke-2006a} and can therefore provide valuable constraints on NS masses and radii.

However, the derived radius constraints are heavily dependent on distance measurements, which are not well known for many LMXBs. Instead, we turn to qLMXBs located in globular clusters, whose distances are known to within $\sim5-10\%$ \citep{Brown-1998,Rutledge-2002a}. With accurate distance measurements, constraints on the NS EOS have been derived from studies of a number of globular cluster qLMXBs. Of the $\sim50$ known qLMXBs in globular clusters, a relatively small fraction have sufficient flux, and low enough extinction for dedicated studies of the NS EOS. These include qLMXBs in 47 Tuc \citep[X5 and X7;][]{Heinke-2003,Heinke-2006a,Bogdanov-2016}, M13 \citep{Gendre-2003b,Webb-2007,Catuneanu-2013}, $\omega$ Cen \citep{Rutledge-2002b,Gendre-2003a,Webb-2007,Heinke-2014}, M28 \citep[source 26;][]{Becker-2003,Servillat-2012}, NGC 6397 \citep[U24;][]{Grindlay-2001,Guillot-2011a,Heinke-2014}, NGC 6304 \citep{Guillot-2009a,Guillot-2009b,Guillot-2013}, NGC 2808 \citep{Webb-2007,Servillat-2008}, NGC 6553 \citep{Guillot-2011b} and M30 \citep{Lugger-2007,Guillot-2014}.

Many previous studies have assumed that the atmosphere of the NS in a qLMXB is purely hydrogen \citep[e.g.][and references therein]{Guillot-2013}. This is a reasonable assumption for typical LMXBs with main sequence donors, as once accretion stops the accreted elements will stratify \citep{Alcock-1980,Romani-1987}. However, it has been noted that between 28 and 44\% of observed bright globular cluster LMXBs are ultracompact sources \citep[i.e. they have orbital periods $<1$hr; see][]{Bahramian-2014a}, suggesting that they require degenerate white dwarf companions devoid of hydrogen, as main sequence stars are not compact enough to exist in such a small orbit. If this fraction transfers to the quiescent population, then it is likely that a significant number of qLMXBs in globular clusters contain NSs with atmospheres composed of heavier elements (He, C, O).

If this is the case, then an X-ray spectrum incorrectly modelled with a hydrogen atmosphere will underestimate the radius, as spectral fits with heavier element atmospheres give larger radii than H atmospheres \citep{Ho-2009}. The atmosphere of the NS therefore has important consequences for the EOS and must be considered \citep[e.g.][]{Servillat-2012,Steiner-2018}. Unfortunately, it is difficult to determine the correct atmosphere to use without utilising optical observations to detect (or not) hydrogen in the optical spectrum of LMXBs \citep[e.g.][]{Haggard-2004, Degenaar-2010a} or identifying the orbital period \citep[e.g.][]{Heinke-2003}. We note here that low-level accretion can work to prevent the stratification of elements \citep{Rutledge-2002a}, which would introduce additional uncertainties in X-ray spectral models. However, the lack of variability in the majority of globular cluster qLMXBs suggests that the accretion is not a dominant process, and therefore the accretion rate is low enough to allow the atmosphere to stratify \citep{Walsh-2015,Bahramian-2015}.

We focus here on the qLMXB source located in the globular cluster M13, discovered by \emph{ROSAT} \citep{Fox-1996,Verbunt-2001} and further studied by the \emph{XMM-Newton} and \emph{Chandra} X-ray observatories \citep{Gendre-2003b,Webb-2007,Servillat-2011,Catuneanu-2013}. Since its discovery, there have been a number of attempts to constrain $R_{\rm NS}$ through X-ray observations. This has resulted in a wide range of measurements, from a relatively compact NS \citep[$R_{\rm NS}\sim9-10$ km][]{Webb-2007,Guillot-2013}, to a much larger one \citep[$R_{\rm NS}\sim12-15$, dependent on the chosen atmosphere;][]{Catuneanu-2013}.

In this paper we utilise a new, deep observation of M13 with \xmm, deriving the tightest constraints on $R_{\rm NS}$ for this NS yet. We target M13 as the NS has a sufficiently high flux and low hydrogen absorption column ($N_{\rm H}$) where a modest \xmm\ observation can make the largest impact in reducing uncertainty on $R_{\rm NS}$. We discuss our results in the context of the NS EOS and comment on the nature of the NS atmosphere in this particular qLMXB.

\section{Observations and Data Reduction}
In this work, we utilize data from \ros, \cha\ and \xmm, focusing in particular on a 2016 $\sim100$ ks observation of M13 with the European Photon Imaging Counter (EPIC) detectors on board \xmm. We also use two \xmm\ observations of the cluster from 2002, a pair of archival \cha\ observations from 2006 and a 1992 \ros\ pointing mode observation (see Table \ref{obs-table}).

The reduction of the \ros\ data and subsequent extraction of the spectrum is described by \citet{Webb-2007} and \citet{Catuneanu-2013}. However, we reprocessed the \cha\ and \xmm\ data using a more recent calibration. The \cha\ data were reduced using \textsc{ciao} (\cha\ Interactive Analysis of Observations) v4.9 and the \cha\ Calibration Database (CALDB) v4.7.3 \citep{Fruscione-2006}. The data were reprocessed with the {\tt chandra\_repro} script to apply the latest calibration updates and bad pixel files. We filtered the data in the energy range 0.3--10 keV and found no evidence for background flaring. The spectra were extracted from circles of radius $2\arcsec$ centred on the qLMXB using the \textsc{ciao} script {\tt specextract} which also generated the corresponding response matrices.

The \xmm\ data were processed with the Science Analysis System (\textsc{sas}) v15.0.0. For all observations, we extracted events from the EPIC pn \citep{Struder-2001} and MOS \citep{Turner-2001} detectors using {\tt epproc} and {\tt emproc}, respectively. All three \xmm~observations revealed signs of background flaring and were therefore filtered to remove the data affected by the periods of the strongest flaring activity. We use filters of 1, 2 and 0.25 count s$^{-1}$ (MOS) and 4.5, 5 and 0.4 count s$^{-1}$ (pn), for the 2002 Jan 28, Jan 30 and 2016 Feb 2 observations, respectively. 

\begin{figure}
	\centering
	\includegraphics[width=0.45\textwidth]{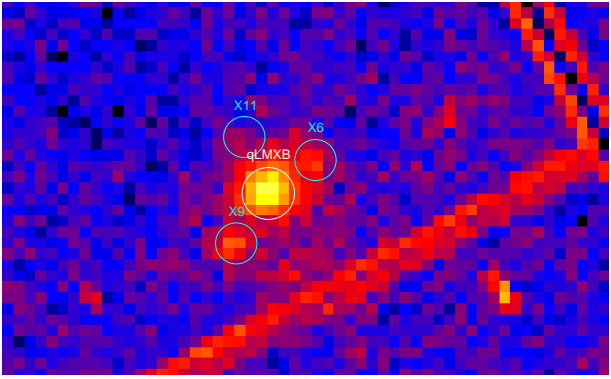}
	\caption{2016 \xmm\ EPIC pn image of the qLMXB in M13. The white circle represents the region used to extract the source spectrum. The cyan circles highlight the \cha\ positions of three nearby X-ray sources, X6, X9 and X11 \citep{Servillat-2011}.}
	\label{regions}
\end{figure}

Circular regions with radii of 9$\arcsec$.5 were used to extract the spectra of the NS qLMXB. This ensured that photons from nearby X-ray sources \citep[X6, X9 and X11;][]{Servillat-2011} were excluded (Fig. \ref{regions}). Response matrices were generated using {\tt rmfgen} and {\tt arfgen} and the spectra were grouped such that they contained at least 20 counts per bin. To achieve better statistics, the two MOS spectra from each observation were combined using the \textsc{heasoft} v6.19 tool {\tt addspec}.\\

\begin{table*}
	\centering
	\caption{X-ray observations of M13.}
	\begin{tabular}{l c c c c}
	\hline
	\hline
	Mission & ObsID & Date & Detector & GTI \\
	& & & & (s)  \\
	\hline
	\ros & RP300181N00 & 1992 Sep & PSPCB & 45872\\
	\xmm & 0085280301 & 2002 Jan 28 & MOS1+MOS2 & 35222  \\
	&&& PN & 14033  \\
	\xmm & 0085280801 & 2002 Jan 30 & MOS1+MOS2 & 30868  \\
	&&& PN & 12032 \\
	\cha & 7290 & 2006 Mar 9 & ACIS-S & 27894  \\
	\cha & 5436 & 2006 Mar 11 & ACIS-S & 26800  \\
	\xmm & 0760750101 & 2016 Feb 2 & MOS1+MOS2 & 96653  \\
	&&& PN & 81587 \\
	\hline
	\label{obs-table}
	\end{tabular}

\end{table*}

\section{Data Analysis and Results}
\subsection{Spectral fits}
All spectral fits were performed using \textsc{xspec} v12.9.1p \citep{Arnaud-1996} which uses the $\chi^2$ minimisation technique to determine the best fit model. The interstellar absorption is accounted for by the {\tt tbabs} model with \citet{Wilms-2000} abundances and photo-ionisation cross-sections described by \citet{Verner-1996}. In all models we assume a distance to M13 of $d=7.7$ kpc \citep{McLaughlin-2005}. We choose this distance as it is consistent with (and between) $d=7.65\pm0.36$ kpc as calculated by \citet{Sandquist-2010} and $d=7.8\pm0.1$ kpc determined by \citet{Recio-Blanco-2005}, as well as to be able to draw comparisons with \citet{Webb-2007} and \citet{Catuneanu-2013}, who both use $d=7.7$ kpc. We used a normalization constant to account for the cross-calibration differences between each detector, allowing the constant to vary relative to the MOS detectors, for which it was fixed to unity. In all fits, in addition to the relevant NS atmosphere model, we also included a thermal bremsstrahlung model for the fit to the \ros~data, to account for the cataclysmic variable (CV) source X6 \citep{Servillat-2011}, which was not resolved as an individual point source by \ros. We fixed $kT$ of the bremsstrahlung model to 4.5 keV \citep{Webb-2007,Catuneanu-2013}. All uncertainties are quoted at the 90\% confidence level, unless otherwise stated.

\subsubsection{Hydrogen atmosphere model}

\begin{figure}
	\centering
	\includegraphics[width=0.45\textwidth]{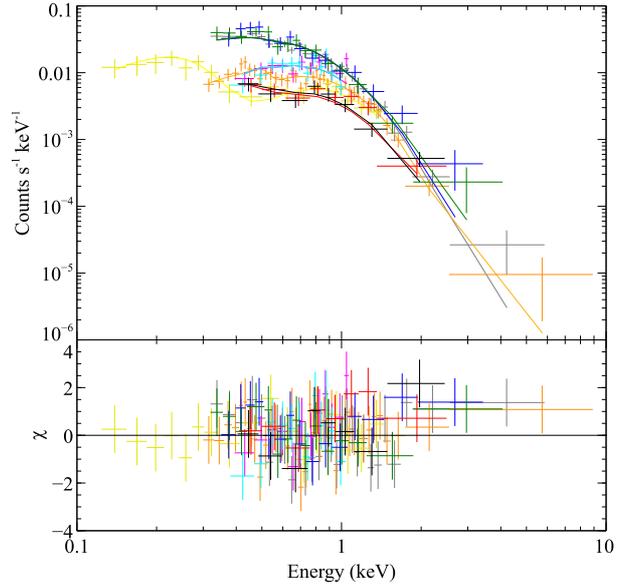}
	\caption{Spectra of the qLMXB in M13, fit with an absorbed hydrogen atmosphere model. Plotted are the 2002 \emph{XMM-Newton} MOS spectra (black and red), 2002 pn spectra (green and blue), 2016 MOS and pn spectra (orange and grey, respectively), 2006 \emph{Chandra} spectra (cyan and magenta) and 1992 \emph{ROSAT} spectrum (yellow). The \ros~data has been fit with an additional bremsstrahlung component to account for the unresolved CV M13 X6. The best-fit model is plotted as a solid line for each spectrum. The bottom panel shows the $\Delta\chi$ residuals.}
	\label{H-spectra}
\end{figure}

\begin{figure}
	\centering
	\includegraphics[width=0.45\textwidth]{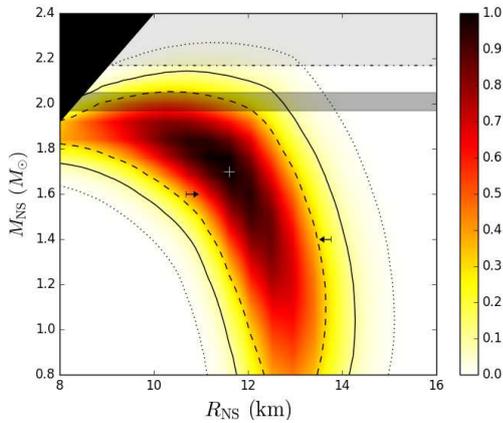}
	\caption{Mass-radius probability distribution for the hydrogen atmosphere model fit to the \xmm, \cha~and \ros~spectra of the qLMXB in M13. The dashed, solid and dotted curves represent the confidence limits at the 68\%, 90\% and 99\% level, respectively. The black shaded section in the upper left region of the plot represents the area forbidden by causality \citep[$R_{\rm NS}<2.82GM/c^{2}$;][]{Haensel-1999}. The light grey shaded region $M_{\rm NS}>2.17{\rm M}_{\odot}$ may be disfavoured based on the interpretation of the NS-NS merger GW170817 \citep[e.g.][]{Margalit-2017}. The narrow, dark grey strip represents the most massive NS measured, PSR J0348+0432 \citep[$M_{\rm NS}=2.01\pm0.04{\rm M}_{\odot}$;][]{Antoniadis-2013}. The arrows represent mass-dependent limits on $R_{\rm NS}$ derived from GW170817 \citep{Fattoyev-2017,Bauswein-2017}.
	} 
	\label{H-contours}
\end{figure}

We fit the \xmm, \cha\ and \ros\ spectra simultaneously with an absorbed hydrogen atmosphere model {\tt nsatmos} \citep{Heinke-2006a}. The spectra are plotted in Fig. \ref{H-spectra} and the best-fit model parameters are displayed in Table \ref{fits}, assuming a fixed NS mass of $M_{\rm NS}=1.4\mathrm{M}_{\odot}$. The resulting absorption column is consistent with $N_{\rm H}=(1.74\pm0.87)\times10^{20}$ cm$^{-2}$ in the direction of M13, which is inferred from the extinction, $E(B-V)=0.02\pm0.01$\footnote{http://physwww.mcmaster.ca/\%7Eharris/mwgc.ref}, derived by \citealt{Harris-1996} (2010 edition), and using \citet{Bahramian-2015} to convert between $A_V$ and $N_{\rm H}$. We determine a best fit NS radius $R_{\mathrm{NS}}=12.2^{+1.5}_{-1.1}$ km. The determined radius is consistent with that of \citet{Catuneanu-2013}, $R_{\mathrm{NS}}=11.7^{+1.9}_{-2.2}$ km, with tighter constraints. Allowing the mass to vary gives a best fit with $M_{\rm NS}=1.7\mathrm{M}_{\odot}$ and $R_{\rm NS}=11.6$ km. To visualise the derived mass-radius relationship we calculate the $\chi^2$ contours with the {\tt steppar} command in \textsc{xspec} and convert this to a probability distribution ${\cal L}\propto\exp(-\chi^2/2)$, shown in Fig. \ref{H-contours} \citep[see e.g.][]{Steiner-2018}.


\subsubsection{Helium atmosphere model}
\begin{figure}
	\centering
	\includegraphics[width=0.45\textwidth]{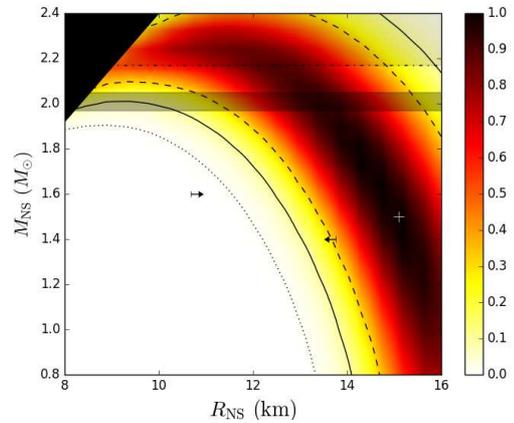}
	\caption{Mass-radius probability distribution for the helium atmosphere model fit to the \xmm, \cha~and \ros~spectra of the qLMXB in M13. The key is the same as in Fig. \ref{H-contours}.}
	\label{He-contours}
\end{figure}

We also fit the spectra with an absorbed helium-atmosphere model {\tt nsx} \citep{Ho-2009}. The best-fit model parameters, assuming a $1.4\mathrm{M}_{\odot}$ NS, are presented in Table \ref{fits}. For a $1.4\mathrm{M}_{\odot}$ NS, the helium atmosphere model determines a best-fit NS radius of $15.1^{+2.0}_{-1.6}$ km, $\sim3$ km larger than that implied by a hydrogen atmosphere. As with the H-atmosphere fits, the inferred absorption column is consistent with that in the direction of M13 \citep[][2010 edition]{Harris-1996}. If we allow the mass to vary we find a best fit $R_{\rm NS}=15.1$ km with a $M_{\rm NS}=1.5{\rm M}_{\odot}$. The mass-radius probability distribution is plotted in Fig. \ref{He-contours}.

\begin{table}
	\centering
	\caption{Best-fit parameters to the \xmm, \cha\ and \ros\ spectra for hydrogen ({\tt nsatmos}) and helium ({\tt nsx}) atmosphere models, with a NS mass fixed to $1.4\mathrm{M}_{\odot}$.}
	\begin{tabular}{l c c}
	\hline
	\hline
	Parameter & {\tt nsatmos} & {\tt nsx} \\
	\hline
	$N_{\mathrm{H}}$ & $0.9^{+0.5}_{-0.4}\times10^{20}$ cm$^{-2}$ & $1.2^{+0.6}_{-0.5}\times10^{20}$ cm$^{-2}$ \\
	$\log_{10}T_{\mathrm{eff}}$ & $5.97\pm0.02$ & $5.92^{+0.02}_{-0.03}$ \\
	$R_{\mathrm{NS}}$ & $12.2^{+1.5}_{-1.1}$ km & $15.1^{+2.0}_{-1.6}$ km \\
	$\chi^2/\mathrm{dof}$ & 128.6/148 & 123.3/148 \\
	\hline
	\label{fits}
	\end{tabular}	
\end{table}

The helium atmosphere model provides a slightly better fit to the spectra ($\Delta\chi^2=5.3$ for the same degrees of freedom) than the hydrogen model. However, both models are considered an acceptable fit to the data. Therefore it is not possible to determine from the fitting which atmosphere model is the correct one.

\section{Discussion}

\subsection{The effect of distance uncertainty}
\begin{figure*}
\captionsetup[subfigure]{labelformat=empty}
\begin{centering}
\subfloat[]{\includegraphics[width=.45\textwidth]{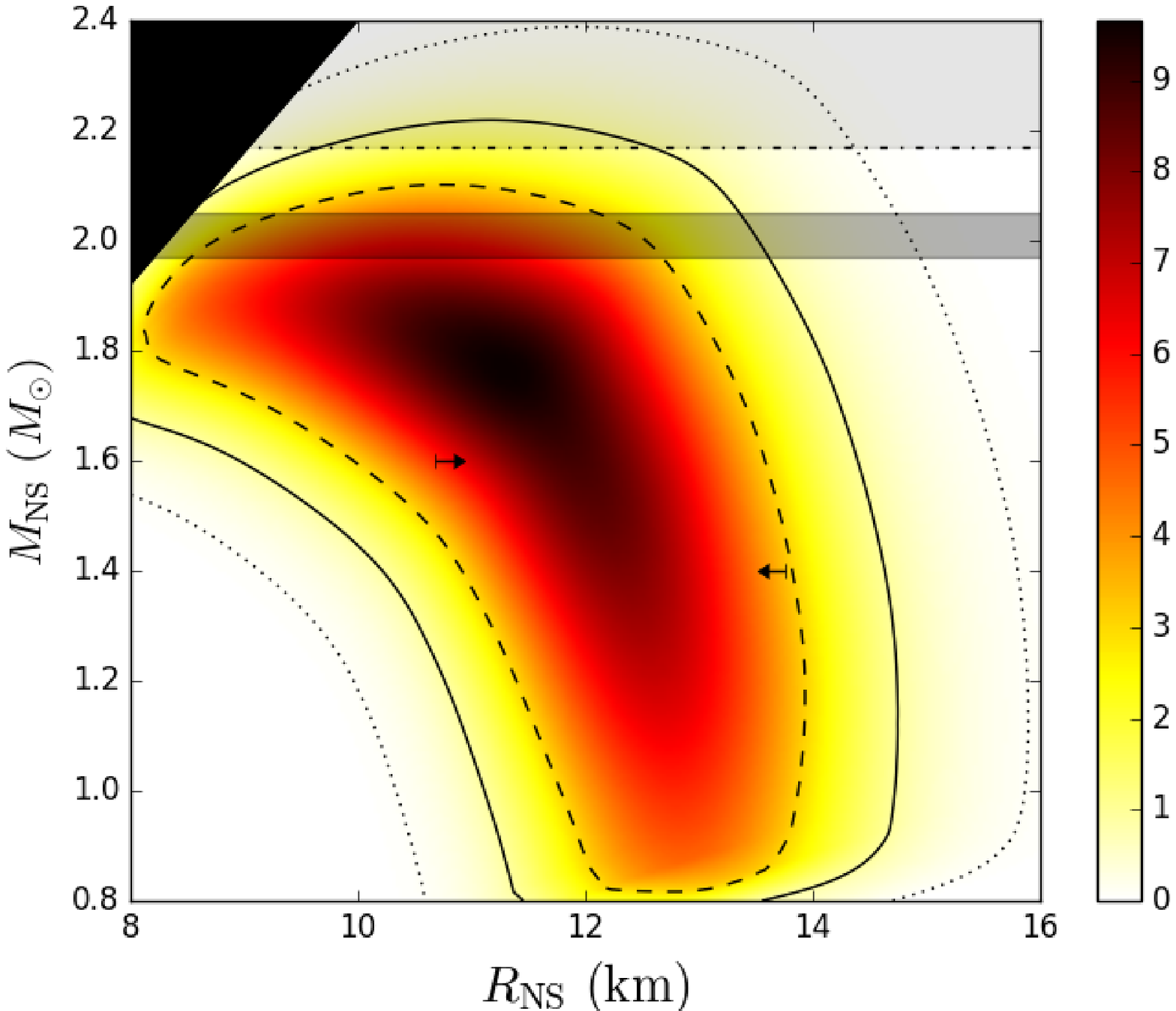}}
\subfloat[]{\includegraphics[width=.45\textwidth]{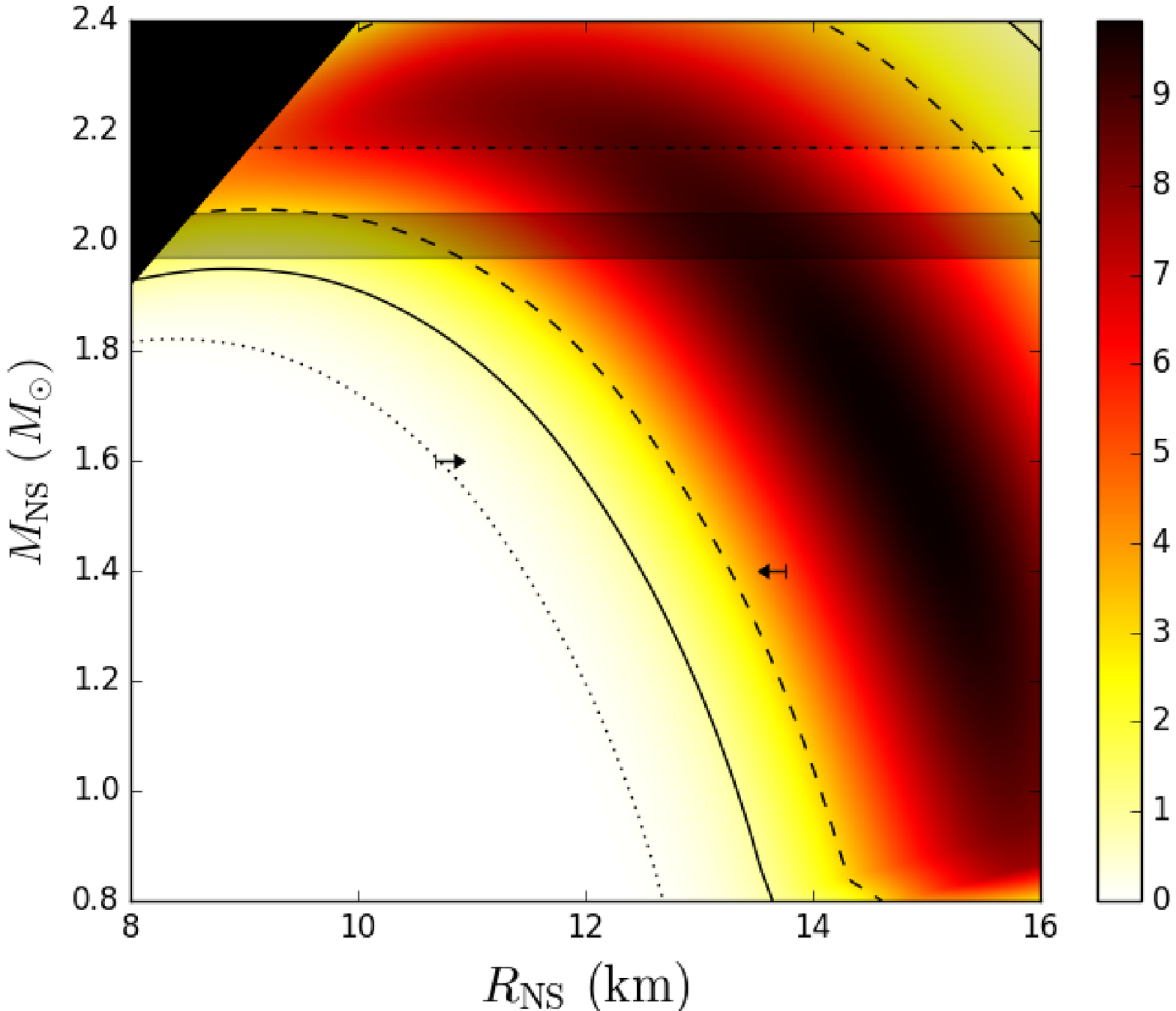}}
\end{centering}
\caption{Mass-radius probability distribution (arbitrary normalisation) for the hydrogen (left) and helium (right) model atmosphere fits to the \xmm, \cha\ and \ros\ spectra of the M13 qLMXB, incorporating an uncertainty in the distance of $\delta d=\pm0.36$ kpc. The key is the same as in Fig. \ref{H-contours}.}
\label{dist_dens}	
\end{figure*}

We have computed our spectral fits assuming a distance to M13 $d=7.7$ kpc \citep{McLaughlin-2005}. However, this did not include its associated uncertainty, which can have an effect on the inferred mass-radius relation. We can incorporate this into the probability distribution by integrating over the distance uncertainty \citep[following the method of][]{Steiner-2018}. Using a conservative uncertainty of $\Delta d=\pm0.36$ kpc \citep{Sandquist-2010}, we find $R_{\rm NS}=12.3^{+1.9}_{-1.7}$ km and $R_{\rm NS}=15.3^{+2.4}_{-2.2}$ km for a $1.4{\rm M}_{\odot}$ NS with a H and He atmosphere, respectively. As expected, introducing an uncertainty on distance increases the radius uncertainties accordingly. The density distributions are plotted in Fig. \ref{dist_dens}, and show a similar increase in the mass-radius contours.

\subsection{Hydrogen vs. helium atmosphere}
We have shown that the NS mass and radius constraints are highly dependent on the chosen atmosphere model. If we fix $M_{\rm NS}=1.4{\rm M}_{\odot}$, we find that $R_{\rm NS}$ increases by $\sim3$ km if the atmosphere is composed of helium versus hydrogen. It is therefore important to distinguish the nature of the companion in order to choose the correct model. \citet{Steiner-2018} model the spectra of eight qLMXBs in globular clusters, and combine them to place constraints on the EOS, based on the knowledge that all NSs must have the same EOS \citep{Lattimer-2001}. \citet{Steiner-2018} calculate the probability of the qLMXB in M13 having a helium atmosphere to be $<28\%$, but with the tighter constraints provided by this work, this probability is likely to be even lower, despite the lower $\chi^2$ for a helium atmosphere model (compared to hydrogen) suggesting a better fit. However, the only reliable way to distinguish between hydrogen and helium atmospheres in NS LMXBs is through direct optical/NIR observations of the counterpart.


\subsection{Neutron star equation of state}

Our results tighten the constraints on the mass and radius of the NS presented by \citet{Catuneanu-2013}, which has implications for the NS EOS. Previous studies of M13 claim tighter constraints on $M_{\rm NS}$ and $R_{\rm NS}$ \citep{Gendre-2003b,Webb-2007} than this work, but these results were not reproduced in later studies \citep{Catuneanu-2013,Guillot-2013}. Our results are consistent with \citet{Catuneanu-2013} and our derived radius (for a $1.4{\rm M}_{\odot}$ NS) falls within the preferred range of 10--14 km for NSs, calculated from observations of multiple qLMXBs and taking into account the effects that a number of uncertainties (e.g. distance, atmosphere composition) have on the inferred mass-radius relation \citep{Steiner-2018}.

A previous study fitting spectra of M13 and other qLMXBs by \citet{Guillot-2013} preferred a smaller $R_{\rm NS}=9.2^{+1.7}_{-2.3p}$km (where $p$ indicates that the parameter was pegged at the hard limit of the model) for M13. However, in that study, a smaller distance was chosen \citep[$d=6.5$ kpc;][]{Rees-1996} rather than the 7.7 kpc we utilise in this work \citep{McLaughlin-2005}, a distance in agreement with measurements by \citet{Recio-Blanco-2005} and \citet{Sandquist-2010} \citep[see also][]{Steiner-2018}. We note that if we instead choose $d=6.5$ kpc, we find $R_{\rm NS}=10.1\pm1.4$km, consistent with the value derived by \citet{Guillot-2013}.

The detection of gravitational waves (GW170817) from two merging NSs \citep{Abbott-2017a} has placed some limits on the NS EOS. The event placed an upper limit on the tidal deformability parameter ($\Lambda$, an intrinsic NS property sensitive to the stellar compactness). Limits on the tidal deformability obtained as the two bodies approached coalescence \citep[$\Lambda\leq800$ for $M_{\rm NS}=1.4{\rm M}_{\odot}$;][]{Abbott-2017a} can be translated into an upper limit on the radius of a $1.4{\rm M}_{\odot}$ NS \citep[$R_{\rm NS}<13.76$ km;][]{Fattoyev-2017}. Information from the short gamma-ray burst that followed the merger 1.7s later \citep[e.g.][]{Abbott-2017b,Goldstein-2017,Savchenko-2017} can be used to infer a limit on the maximum mass of a NS of $\lesssim2.17{\rm M}_{\odot}$ \citep{Margalit-2017, Shibata-2017, Rezzolla-2018}. We must note, however, that these calculations, though all consistent and performed independently of one another, are heavily dependent on physical assumptions about the type of compact object formed in the merger. Finally, \citet{Bauswein-2017} calculated a lower limit of $R_{\rm NS}>10.68$ km (for $M_{\rm NS}=1.6{\rm M}_{\odot}$), also assuming that the merger did not result in a prompt collapse to a black hole - as suggested by the detection of a kilonova \citep{Kasen-2017,Pian-2017,Smartt-2017}. Though the radius constraints inferred from GW170817 do not fully rule out the possibility of a He atmosphere in the M13 qLMXB, they are in agreement with \citet{Steiner-2018} in that it is unlikely to be the case.





\section{Conclusions}

We have derived NS mass-radius constraints for the qLMXB located in the globular cluster M13 using archival observations and a new, deep observation of the cluster with \xmm. We provide the tightest constraints on the radius of the NS ($R_{\mathrm{NS}}=12.2^{+1.5}_{-1.1}$ km assuming a H atmosphere and $M_{\rm NS}=1.4{\rm M}_{\odot}$), which are in good agreement with the limits on the NS EOS derived by \citet{Steiner-2018}. We find that introducing a conservative uncertainty on distance \citep{Sandquist-2010}, increases the radius uncertainties accordingly. We cannot definitively rule out a He atmosphere, but spectral fits infer a much larger $R_{\rm NS}=15.1^{+2.0}_{-1.6}$ km, which overlaps with the upper edge of the $R_{\rm NS}=10-14$ km range derived by \citet{Steiner-2018}. In addition, limits on $R_{\rm NS}$ derived from the NS-NS merger GW170817 \citep{Abbott-2017a,Fattoyev-2017} disfavour a He atmosphere interpretation for the qLMXB in M13. To verify the nature of the atmosphere of the NS, we require spectroscopy of the optical/NIR counterpart, which has not yet been discovered.



\section*{Acknowledgements}
We thank the anonymous referee for useful comments which helped improve the manuscript. AWS would like to thank S. Morsink, G. Sivakoff and R. Fern\'{a}ndez for useful discussions regarding the neutron star equation of state. COH is supported by an NSERC Discovery Grant and a Discovery Accelerator Supplement. AWSteiner is supported by grant NSF PHY 1554876. WCGH acknowledges funding from STFC in the UK through grant number ST/M000931/1.




\bibliographystyle{mnras}
\bibliography{M13_qLMXB.arxiv.bib} 

\begin{thebibliography}{}
\makeatletter
\relax
\def\mn@urlcharsother{\let\do\@makeother \do\$\do\&\do\#\do\^\do\_\do\%\do\~}
\def\mn@doi{\begingroup\mn@urlcharsother \@ifnextchar [ {\mn@doi@}
  {\mn@doi@[]}}
\def\mn@doi@[#1]#2{\def\@tempa{#1}\ifx\@tempa\@empty \href
  {http://dx.doi.org/#2} {doi:#2}\else \href {http://dx.doi.org/#2} {#1}\fi
  \endgroup}
\def\mn@eprint#1#2{\mn@eprint@#1:#2::\@nil}
\def\mn@eprint@arXiv#1{\href {http://arxiv.org/abs/#1} {{\tt arXiv:#1}}}
\def\mn@eprint@dblp#1{\href {http://dblp.uni-trier.de/rec/bibtex/#1.xml}
  {dblp:#1}}
\def\mn@eprint@#1:#2:#3:#4\@nil{\def\@tempa {#1}\def\@tempb {#2}\def\@tempc
  {#3}\ifx \@tempc \@empty \let \@tempc \@tempb \let \@tempb \@tempa \fi \ifx
  \@tempb \@empty \def\@tempb {arXiv}\fi \@ifundefined
  {mn@eprint@\@tempb}{\@tempb:\@tempc}{\expandafter \expandafter \csname
  mn@eprint@\@tempb\endcsname \expandafter{\@tempc}}}

\bibitem[\protect\citeauthoryear{{Abbott} et~al.,}{{Abbott}
  et~al.}{2017a}]{Abbott-2017a}
{Abbott} B.~P.,  et~al., 2017a, \mn@doi [Physical Review Letters]
  {10.1103/PhysRevLett.119.161101}, \href
  {http://adsabs.harvard.edu/abs/2017PhRvL.119p1101A} {119, 161101}

\bibitem[\protect\citeauthoryear{{Abbott} et~al.,}{{Abbott}
  et~al.}{2017b}]{Abbott-2017b}
{Abbott} B.~P.,  et~al., 2017b, \mn@doi [\apjl] {10.3847/2041-8213/aa91c9},
  \href {http://adsabs.harvard.edu/abs/2017ApJ...848L..12A} {848, L12}

\bibitem[\protect\citeauthoryear{{Alcock} \& {Illarionov}}{{Alcock} \&
  {Illarionov}}{1980}]{Alcock-1980}
{Alcock} C.,  {Illarionov} A.,  1980, \mn@doi [\apj] {10.1086/157656}, \href
  {http://adsabs.harvard.edu/abs/1980ApJ...235..534A} {235, 534}

\bibitem[\protect\citeauthoryear{{Antoniadis} et~al.,}{{Antoniadis}
  et~al.}{2013}]{Antoniadis-2013}
{Antoniadis} J.,  et~al., 2013, \mn@doi [Science] {10.1126/science.1233232},
  \href {http://adsabs.harvard.edu/abs/2013Sci...340..448A} {340, 448}

\bibitem[\protect\citeauthoryear{{Arnaud}}{{Arnaud}}{1996}]{Arnaud-1996}
{Arnaud} K.~A.,  1996, in {Jacoby} G.~H.,  {Barnes} J.,  eds,  Astronomical
  Society of the Pacific Conference Series Vol. 101, Astronomical Data Analysis
  Software and Systems V. p.~17

\bibitem[\protect\citeauthoryear{{Bahramian} et~al.,}{{Bahramian}
  et~al.}{2014}]{Bahramian-2014a}
{Bahramian} A.,  et~al., 2014, \mn@doi [\apj] {10.1088/0004-637X/780/2/127},
  \href {http://adsabs.harvard.edu/abs/2014ApJ...780..127B} {780, 127}

\bibitem[\protect\citeauthoryear{{Bahramian}, {Heinke}, {Degenaar}, {Chomiuk},
  {Wijnands}, {Strader}, {Ho}  \& {Pooley}}{{Bahramian}
  et~al.}{2015}]{Bahramian-2015}
{Bahramian} A.,  {Heinke} C.~O.,  {Degenaar} N.,  {Chomiuk} L.,  {Wijnands} R.,
   {Strader} J.,  {Ho} W.~C.~G.,   {Pooley} D.,  2015, \mn@doi [\mnras]
  {10.1093/mnras/stv1585}, \href
  {http://adsabs.harvard.edu/abs/2015MNRAS.452.3475B} {452, 3475}

\bibitem[\protect\citeauthoryear{{Bauswein}, {Just}, {Janka}  \&
  {Stergioulas}}{{Bauswein} et~al.}{2017}]{Bauswein-2017}
{Bauswein} A.,  {Just} O.,  {Janka} H.-T.,   {Stergioulas} N.,  2017, \mn@doi
  [\apjl] {10.3847/2041-8213/aa9994}, \href
  {http://adsabs.harvard.edu/abs/2017ApJ...850L..34B} {850, L34}

\bibitem[\protect\citeauthoryear{{Becker} et~al.,}{{Becker}
  et~al.}{2003}]{Becker-2003}
{Becker} W.,  et~al., 2003, \mn@doi [\apj] {10.1086/376967}, \href
  {http://adsabs.harvard.edu/abs/2003ApJ...594..798B} {594, 798}

\bibitem[\protect\citeauthoryear{{Bogdanov}, {Heinke}, {{\"O}zel}  \&
  {G{\"u}ver}}{{Bogdanov} et~al.}{2016}]{Bogdanov-2016}
{Bogdanov} S.,  {Heinke} C.~O.,  {{\"O}zel} F.,   {G{\"u}ver} T.,  2016,
  \mn@doi [\apj] {10.3847/0004-637X/831/2/184}, \href
  {http://adsabs.harvard.edu/abs/2016ApJ...831..184B} {831, 184}

\bibitem[\protect\citeauthoryear{{Brown}, {Bildsten}  \& {Rutledge}}{{Brown}
  et~al.}{1998}]{Brown-1998}
{Brown} E.~F.,  {Bildsten} L.,   {Rutledge} R.~E.,  1998, \mn@doi [\apjl]
  {10.1086/311578}, \href {http://adsabs.harvard.edu/abs/1998ApJ...504L..95B}
  {504, L95}

\bibitem[\protect\citeauthoryear{{Campana}, {Colpi}, {Mereghetti}, {Stella}  \&
  {Tavani}}{{Campana} et~al.}{1998}]{Campana-1998}
{Campana} S.,  {Colpi} M.,  {Mereghetti} S.,  {Stella} L.,   {Tavani} M.,
  1998, \mn@doi [\aapr] {10.1007/s001590050012}, \href
  {http://adsabs.harvard.edu/abs/1998A%26ARv...8..279C} {8, 279}

\bibitem[\protect\citeauthoryear{{Catuneanu}, {Heinke}, {Sivakoff}, {Ho}  \&
  {Servillat}}{{Catuneanu} et~al.}{2013}]{Catuneanu-2013}
{Catuneanu} A.,  {Heinke} C.~O.,  {Sivakoff} G.~R.,  {Ho} W.~C.~G.,
  {Servillat} M.,  2013, \mn@doi [\apj] {10.1088/0004-637X/764/2/145}, \href
  {http://adsabs.harvard.edu/abs/2013ApJ...764..145C} {764, 145}

\bibitem[\protect\citeauthoryear{{Degenaar} et~al.,}{{Degenaar}
  et~al.}{2010}]{Degenaar-2010a}
{Degenaar} N.,  et~al., 2010, \mn@doi [\mnras]
  {10.1111/j.1365-2966.2010.16388.x}, \href
  {http://adsabs.harvard.edu/abs/2010MNRAS.404.1591D} {404, 1591}

\bibitem[\protect\citeauthoryear{{Degenaar}, {Brown}  \& {Wijnands}}{{Degenaar}
  et~al.}{2011}]{Degenaar-2011b}
{Degenaar} N.,  {Brown} E.~F.,   {Wijnands} R.,  2011, \mn@doi [\mnras]
  {10.1111/j.1745-3933.2011.01164.x}, \href
  {http://adsabs.harvard.edu/abs/2011MNRAS.418L.152D} {418, L152}

\bibitem[\protect\citeauthoryear{{Fattoyev}, {Piekarewicz}  \&
  {Horowitz}}{{Fattoyev} et~al.}{2017}]{Fattoyev-2017}
{Fattoyev} F.~J.,  {Piekarewicz} J.,   {Horowitz} C.~J.,  2017, preprint, \href
  {http://adsabs.harvard.edu/abs/2017arXiv171106615F} {} (\mn@eprint {arXiv}
  {1711.06615})

\bibitem[\protect\citeauthoryear{{Fox}, {Lewin}, {Margon}, {van Paradijs}  \&
  {Verbunt}}{{Fox} et~al.}{1996}]{Fox-1996}
{Fox} D.,  {Lewin} W.,  {Margon} B.,  {van Paradijs} J.,   {Verbunt} F.,  1996,
  \mn@doi [\mnras] {10.1093/mnras/282.3.1027}, \href
  {http://adsabs.harvard.edu/abs/1996MNRAS.282.1027F} {282, 1027}

\bibitem[\protect\citeauthoryear{{Fruscione} et~al.,}{{Fruscione}
  et~al.}{2006}]{Fruscione-2006}
{Fruscione} A.,  et~al., 2006, in Society of Photo-Optical Instrumentation
  Engineers (SPIE) Conference Series. p. 62701V, \mn@doi{10.1117/12.671760}

\bibitem[\protect\citeauthoryear{{Gendre}, {Barret}  \& {Webb}}{{Gendre}
  et~al.}{2003a}]{Gendre-2003a}
{Gendre} B.,  {Barret} D.,   {Webb} N.~A.,  2003a, \mn@doi [\aap]
  {10.1051/0004-6361:20021845}, \href
  {http://adsabs.harvard.edu/abs/2003A%26A...400..521G} {400, 521}

\bibitem[\protect\citeauthoryear{{Gendre}, {Barret}  \& {Webb}}{{Gendre}
  et~al.}{2003b}]{Gendre-2003b}
{Gendre} B.,  {Barret} D.,   {Webb} N.,  2003b, \mn@doi [\aap]
  {10.1051/0004-6361:20030423}, \href
  {http://adsabs.harvard.edu/abs/2003A%26A...403L..11G} {403, L11}

\bibitem[\protect\citeauthoryear{{Goldstein} et~al.,}{{Goldstein}
  et~al.}{2017}]{Goldstein-2017}
{Goldstein} A.,  et~al., 2017, \mn@doi [\apjl] {10.3847/2041-8213/aa8f41},
  \href {http://adsabs.harvard.edu/abs/2017ApJ...848L..14G} {848, L14}

\bibitem[\protect\citeauthoryear{{Grindlay}, {Heinke}, {Edmonds}, {Murray}  \&
  {Cool}}{{Grindlay} et~al.}{2001}]{Grindlay-2001}
{Grindlay} J.~E.,  {Heinke} C.~O.,  {Edmonds} P.~D.,  {Murray} S.~S.,   {Cool}
  A.~M.,  2001, \mn@doi [\apjl] {10.1086/338499}, \href
  {http://adsabs.harvard.edu/abs/2001ApJ...563L..53G} {563, L53}

\bibitem[\protect\citeauthoryear{{Guillot} \& {Rutledge}}{{Guillot} \&
  {Rutledge}}{2014}]{Guillot-2014}
{Guillot} S.,  {Rutledge} R.~E.,  2014, \mn@doi [\apjl]
  {10.1088/2041-8205/796/1/L3}, \href
  {http://adsabs.harvard.edu/abs/2014ApJ...796L...3G} {796, L3}

\bibitem[\protect\citeauthoryear{{Guillot}, {Rutledge}, {Bildsten}, {Brown},
  {Pavlov}  \& {Zavlin}}{{Guillot} et~al.}{2009a}]{Guillot-2009a}
{Guillot} S.,  {Rutledge} R.~E.,  {Bildsten} L.,  {Brown} E.~F.,  {Pavlov}
  G.~G.,   {Zavlin} V.~E.,  2009a, \mn@doi [\mnras]
  {10.1111/j.1365-2966.2008.14076.x}, \href
  {http://adsabs.harvard.edu/abs/2009MNRAS.392..665G} {392, 665}

\bibitem[\protect\citeauthoryear{{Guillot}, {Rutledge}, {Brown}, {Pavlov}  \&
  {Zavlin}}{{Guillot} et~al.}{2009b}]{Guillot-2009b}
{Guillot} S.,  {Rutledge} R.~E.,  {Brown} E.~F.,  {Pavlov} G.~G.,   {Zavlin}
  V.~E.,  2009b, \mn@doi [\apj] {10.1088/0004-637X/699/2/1418}, \href
  {http://adsabs.harvard.edu/abs/2009ApJ...699.1418G} {699, 1418}

\bibitem[\protect\citeauthoryear{{Guillot}, {Rutledge}  \& {Brown}}{{Guillot}
  et~al.}{2011a}]{Guillot-2011a}
{Guillot} S.,  {Rutledge} R.~E.,   {Brown} E.~F.,  2011a, \mn@doi [\apj]
  {10.1088/0004-637X/732/2/88}, \href
  {http://adsabs.harvard.edu/abs/2011ApJ...732...88G} {732, 88}

\bibitem[\protect\citeauthoryear{{Guillot}, {Rutledge}, {Brown}, {Pavlov}  \&
  {Zavlin}}{{Guillot} et~al.}{2011b}]{Guillot-2011b}
{Guillot} S.,  {Rutledge} R.~E.,  {Brown} E.~F.,  {Pavlov} G.~G.,   {Zavlin}
  V.~E.,  2011b, \mn@doi [\apj] {10.1088/0004-637X/738/2/129}, \href
  {http://adsabs.harvard.edu/abs/2011ApJ...738..129G} {738, 129}

\bibitem[\protect\citeauthoryear{{Guillot}, {Servillat}, {Webb}  \&
  {Rutledge}}{{Guillot} et~al.}{2013}]{Guillot-2013}
{Guillot} S.,  {Servillat} M.,  {Webb} N.~A.,   {Rutledge} R.~E.,  2013,
  \mn@doi [\apj] {10.1088/0004-637X/772/1/7}, \href
  {http://adsabs.harvard.edu/abs/2013ApJ...772....7G} {772, 7}

\bibitem[\protect\citeauthoryear{{Haensel}, {Lasota}  \& {Zdunik}}{{Haensel}
  et~al.}{1999}]{Haensel-1999}
{Haensel} P.,  {Lasota} J.~P.,   {Zdunik} J.~L.,  1999, \aap, \href
  {http://adsabs.harvard.edu/abs/1999A%26A...344..151H} {344, 151}

\bibitem[\protect\citeauthoryear{{Haggard}, {Cool}, {Anderson}, {Edmonds},
  {Callanan}, {Heinke}, {Grindlay}  \& {Bailyn}}{{Haggard}
  et~al.}{2004}]{Haggard-2004}
{Haggard} D.,  {Cool} A.~M.,  {Anderson} J.,  {Edmonds} P.~D.,  {Callanan}
  P.~J.,  {Heinke} C.~O.,  {Grindlay} J.~E.,   {Bailyn} C.~D.,  2004, \mn@doi
  [\apj] {10.1086/421549}, \href
  {http://adsabs.harvard.edu/abs/2004ApJ...613..512H} {613, 512}

\bibitem[\protect\citeauthoryear{{Harris}}{{Harris}}{1996}]{Harris-1996}
{Harris} W.~E.,  1996, \mn@doi [\aj] {10.1086/118116}, \href
  {http://adsabs.harvard.edu/abs/1996AJ....112.1487H} {112, 1487}

\bibitem[\protect\citeauthoryear{{Heinke}, {Grindlay}, {Lloyd}  \&
  {Edmonds}}{{Heinke} et~al.}{2003}]{Heinke-2003}
{Heinke} C.~O.,  {Grindlay} J.~E.,  {Lloyd} D.~A.,   {Edmonds} P.~D.,  2003,
  \mn@doi [\apj] {10.1086/374039}, \href
  {http://adsabs.harvard.edu/abs/2003ApJ...588..452H} {588, 452}

\bibitem[\protect\citeauthoryear{{Heinke}, {Rybicki}, {Narayan}  \&
  {Grindlay}}{{Heinke} et~al.}{2006}]{Heinke-2006a}
{Heinke} C.~O.,  {Rybicki} G.~B.,  {Narayan} R.,   {Grindlay} J.~E.,  2006,
  \mn@doi [\apj] {10.1086/503701}, \href
  {http://adsabs.harvard.edu/abs/2006ApJ...644.1090H} {644, 1090}

\bibitem[\protect\citeauthoryear{{Heinke} et~al.,}{{Heinke}
  et~al.}{2014}]{Heinke-2014}
{Heinke} C.~O.,  et~al., 2014, \mn@doi [\mnras] {10.1093/mnras/stu1449}, \href
  {http://adsabs.harvard.edu/abs/2014MNRAS.444..443H} {444, 443}

\bibitem[\protect\citeauthoryear{{Ho} \& {Heinke}}{{Ho} \&
  {Heinke}}{2009}]{Ho-2009}
{Ho} W.~C.~G.,  {Heinke} C.~O.,  2009, \mn@doi [\nat] {10.1038/nature08525},
  \href {http://adsabs.harvard.edu/abs/2009Natur.462...71H} {462, 71}

\bibitem[\protect\citeauthoryear{{Kasen}, {Metzger}, {Barnes}, {Quataert}  \&
  {Ramirez-Ruiz}}{{Kasen} et~al.}{2017}]{Kasen-2017}
{Kasen} D.,  {Metzger} B.,  {Barnes} J.,  {Quataert} E.,   {Ramirez-Ruiz} E.,
  2017, \mn@doi [\nat] {10.1038/nature24453}, \href
  {http://adsabs.harvard.edu/abs/2017Natur.551...80K} {551, 80}

\bibitem[\protect\citeauthoryear{{Lattimer} \& {Prakash}}{{Lattimer} \&
  {Prakash}}{2001}]{Lattimer-2001}
{Lattimer} J.~M.,  {Prakash} M.,  2001, \mn@doi [\apj] {10.1086/319702}, \href
  {http://adsabs.harvard.edu/abs/2001ApJ...550..426L} {550, 426}

\bibitem[\protect\citeauthoryear{{Lugger}, {Cohn}, {Heinke}, {Grindlay}  \&
  {Edmonds}}{{Lugger} et~al.}{2007}]{Lugger-2007}
{Lugger} P.~M.,  {Cohn} H.~N.,  {Heinke} C.~O.,  {Grindlay} J.~E.,   {Edmonds}
  P.~D.,  2007, \mn@doi [\apj] {10.1086/507572}, \href
  {http://adsabs.harvard.edu/abs/2007ApJ...657..286L} {657, 286}

\bibitem[\protect\citeauthoryear{{Margalit} \& {Metzger}}{{Margalit} \&
  {Metzger}}{2017}]{Margalit-2017}
{Margalit} B.,  {Metzger} B.~D.,  2017, \mn@doi [\apjl]
  {10.3847/2041-8213/aa991c}, \href
  {http://adsabs.harvard.edu/abs/2017ApJ...850L..19M} {850, L19}

\bibitem[\protect\citeauthoryear{{McLaughlin} \& {van der Marel}}{{McLaughlin}
  \& {van der Marel}}{2005}]{McLaughlin-2005}
{McLaughlin} D.~E.,  {van der Marel} R.~P.,  2005, \mn@doi [\apjs]
  {10.1086/497429}, \href {http://adsabs.harvard.edu/abs/2005ApJS..161..304M}
  {161, 304}

\bibitem[\protect\citeauthoryear{{N{\"a}ttil{\"a}}, {Miller}, {Steiner},
  {Kajava}, {Suleimanov}  \& {Poutanen}}{{N{\"a}ttil{\"a}}
  et~al.}{2017}]{Nattila-2017}
{N{\"a}ttil{\"a}} J.,  {Miller} M.~C.,  {Steiner} A.~W.,  {Kajava} J.~J.~E.,
  {Suleimanov} V.~F.,   {Poutanen} J.,  2017, \mn@doi [\aap]
  {10.1051/0004-6361/201731082}, \href
  {http://adsabs.harvard.edu/abs/2017A%26A...608A..31N} {608, A31}

\bibitem[\protect\citeauthoryear{{{\"O}zel}, {G{\"u}ver}  \&
  {Psaltis}}{{{\"O}zel} et~al.}{2009}]{Ozel-2009}
{{\"O}zel} F.,  {G{\"u}ver} T.,   {Psaltis} D.,  2009, \mn@doi [\apj]
  {10.1088/0004-637X/693/2/1775}, \href
  {http://adsabs.harvard.edu/abs/2009ApJ...693.1775O} {693, 1775}

\bibitem[\protect\citeauthoryear{{{\"O}zel}, {Gould}  \&
  {G{\"u}ver}}{{{\"O}zel} et~al.}{2012}]{Ozel-2012a}
{{\"O}zel} F.,  {Gould} A.,   {G{\"u}ver} T.,  2012, \mn@doi [\apj]
  {10.1088/0004-637X/748/1/5}, \href
  {http://adsabs.harvard.edu/abs/2012ApJ...748....5O} {748, 5}

\bibitem[\protect\citeauthoryear{{{\"O}zel}, {Psaltis}, {G{\"u}ver}, {Baym},
  {Heinke}  \& {Guillot}}{{{\"O}zel} et~al.}{2016a}]{Ozel-2016a}
{{\"O}zel} F.,  {Psaltis} D.,  {G{\"u}ver} T.,  {Baym} G.,  {Heinke} C.,
  {Guillot} S.,  2016a, \mn@doi [\apj] {10.3847/0004-637X/820/1/28}, \href
  {http://adsabs.harvard.edu/abs/2016ApJ...820...28O} {820, 28}

\bibitem[\protect\citeauthoryear{{{\"O}zel}, {Psaltis}, {Arzoumanian},
  {Morsink}  \& {Baub{\"o}ck}}{{{\"O}zel} et~al.}{2016b}]{Ozel-2016b}
{{\"O}zel} F.,  {Psaltis} D.,  {Arzoumanian} Z.,  {Morsink} S.,   {Baub{\"o}ck}
  M.,  2016b, \mn@doi [\apj] {10.3847/0004-637X/832/1/92}, \href
  {http://adsabs.harvard.edu/abs/2016ApJ...832...92O} {832, 92}

\bibitem[\protect\citeauthoryear{{Pian} et~al.,}{{Pian}
  et~al.}{2017}]{Pian-2017}
{Pian} E.,  et~al., 2017, \mn@doi [\nat] {10.1038/nature24298}, \href
  {http://adsabs.harvard.edu/abs/2017Natur.551...67P} {551, 67}

\bibitem[\protect\citeauthoryear{{Poutanen}, {N{\"a}ttil{\"a}}, {Kajava},
  {Latvala}, {Galloway}, {Kuulkers}  \& {Suleimanov}}{{Poutanen}
  et~al.}{2014}]{Poutanen-2014}
{Poutanen} J.,  {N{\"a}ttil{\"a}} J.,  {Kajava} J.~J.~E.,  {Latvala} O.-M.,
  {Galloway} D.~K.,  {Kuulkers} E.,   {Suleimanov} V.~F.,  2014, \mn@doi
  [\mnras] {10.1093/mnras/stu1139}, \href
  {http://adsabs.harvard.edu/abs/2014MNRAS.442.3777P} {442, 3777}

\bibitem[\protect\citeauthoryear{{Recio-Blanco} et~al.,}{{Recio-Blanco}
  et~al.}{2005}]{Recio-Blanco-2005}
{Recio-Blanco} A.,  et~al., 2005, \mn@doi [\aap] {10.1051/0004-6361:20041053},
  \href {http://adsabs.harvard.edu/abs/2005A%26A...432..851R} {432, 851}

\bibitem[\protect\citeauthoryear{{Rees}}{{Rees}}{1996}]{Rees-1996}
{Rees} Jr. R.~F.,  1996, in {Morrison} H.~L.,  {Sarajedini} A.,  eds,
  Astronomical Society of the Pacific Conference Series Vol. 92, Formation of
  the Galactic Halo...Inside and Out. p.~289

\bibitem[\protect\citeauthoryear{{Rezzolla}, {Most}  \& {Weih}}{{Rezzolla}
  et~al.}{2018}]{Rezzolla-2018}
{Rezzolla} L.,  {Most} E.~R.,   {Weih} L.~R.,  2018, \mn@doi [\apjl]
  {10.3847/2041-8213/aaa401}, \href
  {http://adsabs.harvard.edu/abs/2018ApJ...852L..25R} {852, L25}

\bibitem[\protect\citeauthoryear{{Romani}}{{Romani}}{1987}]{Romani-1987}
{Romani} R.~W.,  1987, \mn@doi [\apj] {10.1086/165010}, \href
  {http://adsabs.harvard.edu/abs/1987ApJ...313..718R} {313, 718}

\bibitem[\protect\citeauthoryear{{Rutledge}, {Bildsten}, {Brown}, {Pavlov}  \&
  {Zavlin}}{{Rutledge} et~al.}{2001a}]{Rutledge-2001a}
{Rutledge} R.~E.,  {Bildsten} L.,  {Brown} E.~F.,  {Pavlov} G.~G.,   {Zavlin}
  V.~E.,  2001a, \mn@doi [\apj] {10.1086/320247}, \href
  {http://adsabs.harvard.edu/abs/2001ApJ...551..921R} {551, 921}

\bibitem[\protect\citeauthoryear{{Rutledge}, {Bildsten}, {Brown}, {Pavlov}  \&
  {Zavlin}}{{Rutledge} et~al.}{2001b}]{Rutledge-2001b}
{Rutledge} R.~E.,  {Bildsten} L.,  {Brown} E.~F.,  {Pavlov} G.~G.,   {Zavlin}
  V.~E.,  2001b, \mn@doi [\apj] {10.1086/322361}, \href
  {http://adsabs.harvard.edu/abs/2001ApJ...559.1054R} {559, 1054}

\bibitem[\protect\citeauthoryear{{Rutledge}, {Bildsten}, {Brown}, {Pavlov}  \&
  {Zavlin}}{{Rutledge} et~al.}{2002a}]{Rutledge-2002a}
{Rutledge} R.~E.,  {Bildsten} L.,  {Brown} E.~F.,  {Pavlov} G.~G.,   {Zavlin}
  V.~E.,  2002a, \mn@doi [\apj] {10.1086/342155}, \href
  {http://adsabs.harvard.edu/abs/2002ApJ...577..346R} {577, 346}

\bibitem[\protect\citeauthoryear{{Rutledge}, {Bildsten}, {Brown}, {Pavlov}  \&
  {Zavlin}}{{Rutledge} et~al.}{2002b}]{Rutledge-2002b}
{Rutledge} R.~E.,  {Bildsten} L.,  {Brown} E.~F.,  {Pavlov} G.~G.,   {Zavlin}
  V.~E.,  2002b, \mn@doi [\apj] {10.1086/342306}, \href
  {http://adsabs.harvard.edu/abs/2002ApJ...578..405R} {578, 405}

\bibitem[\protect\citeauthoryear{{Sandquist}, {Gordon}, {Levine}  \&
  {Bolte}}{{Sandquist} et~al.}{2010}]{Sandquist-2010}
{Sandquist} E.~L.,  {Gordon} M.,  {Levine} D.,   {Bolte} M.,  2010, \mn@doi
  [\aj] {10.1088/0004-6256/139/6/2374}, \href
  {http://adsabs.harvard.edu/abs/2010AJ....139.2374S} {139, 2374}

\bibitem[\protect\citeauthoryear{{Savchenko} et~al.,}{{Savchenko}
  et~al.}{2017}]{Savchenko-2017}
{Savchenko} V.,  et~al., 2017, \mn@doi [\apjl] {10.3847/2041-8213/aa8f94},
  \href {http://adsabs.harvard.edu/abs/2017ApJ...848L..15S} {848, L15}

\bibitem[\protect\citeauthoryear{{Servillat} et~al.,}{{Servillat}
  et~al.}{2008}]{Servillat-2008}
{Servillat} M.,  et~al., 2008, \mn@doi [\aap] {10.1051/0004-6361:200810188},
  \href {http://adsabs.harvard.edu/abs/2008A%26A...490..641S} {490, 641}

\bibitem[\protect\citeauthoryear{{Servillat}, {Webb}, {Lewis}, {Knigge}, {van
  den Berg}, {Dieball}  \& {Grindlay}}{{Servillat}
  et~al.}{2011}]{Servillat-2011}
{Servillat} M.,  {Webb} N.~A.,  {Lewis} F.,  {Knigge} C.,  {van den Berg} M.,
  {Dieball} A.,   {Grindlay} J.,  2011, \mn@doi [\apj]
  {10.1088/0004-637X/733/2/106}, \href
  {http://adsabs.harvard.edu/abs/2011ApJ...733..106S} {733, 106}

\bibitem[\protect\citeauthoryear{{Servillat}, {Heinke}, {Ho}, {Grindlay},
  {Hong}, {van den Berg}  \& {Bogdanov}}{{Servillat}
  et~al.}{2012}]{Servillat-2012}
{Servillat} M.,  {Heinke} C.~O.,  {Ho} W.~C.~G.,  {Grindlay} J.~E.,  {Hong} J.,
   {van den Berg} M.,   {Bogdanov} S.,  2012, \mn@doi [\mnras]
  {10.1111/j.1365-2966.2012.20976.x}, \href
  {http://adsabs.harvard.edu/abs/2012MNRAS.423.1556S} {423, 1556}

\bibitem[\protect\citeauthoryear{{Shibata}, {Fujibayashi}, {Hotokezaka},
  {Kiuchi}, {Kyutoku}, {Sekiguchi}  \& {Tanaka}}{{Shibata}
  et~al.}{2017}]{Shibata-2017}
{Shibata} M.,  {Fujibayashi} S.,  {Hotokezaka} K.,  {Kiuchi} K.,  {Kyutoku} K.,
   {Sekiguchi} Y.,   {Tanaka} M.,  2017, \mn@doi [\prd]
  {10.1103/PhysRevD.96.123012}, \href
  {http://adsabs.harvard.edu/abs/2017PhRvD..96l3012S} {96, 123012}

\bibitem[\protect\citeauthoryear{{Smartt} et~al.,}{{Smartt}
  et~al.}{2017}]{Smartt-2017}
{Smartt} S.~J.,  et~al., 2017, \mn@doi [\nat] {10.1038/nature24303}, \href
  {http://adsabs.harvard.edu/abs/2017Natur.551...75S} {551, 75}

\bibitem[\protect\citeauthoryear{{Steiner}, {Heinke}, {Bogdanov}, {Li}, {Ho},
  {Bahramian}  \& {Han}}{{Steiner} et~al.}{2018}]{Steiner-2018}
{Steiner} A.~W.,  {Heinke} C.~O.,  {Bogdanov} S.,  {Li} C.,  {Ho} W.~C.~G.,
  {Bahramian} A.,   {Han} S.,  2018, \mn@doi [\mnras] {10.1093/mnras/sty215},
  \href {http://adsabs.harvard.edu/abs/2018MNRAS.tmp..216S} {}

\bibitem[\protect\citeauthoryear{{Str{\"u}der} et~al.,}{{Str{\"u}der}
  et~al.}{2001}]{Struder-2001}
{Str{\"u}der} L.,  et~al., 2001, \mn@doi [\aap] {10.1051/0004-6361:20000066},
  \href {http://adsabs.harvard.edu/abs/2001A%26A...365L..18S} {365, L18}

\bibitem[\protect\citeauthoryear{{Turner} et~al.,}{{Turner}
  et~al.}{2001}]{Turner-2001}
{Turner} M.~J.~L.,  et~al., 2001, \mn@doi [\aap] {10.1051/0004-6361:20000087},
  \href {http://adsabs.harvard.edu/abs/2001A%26A...365L..27T} {365, L27}

\bibitem[\protect\citeauthoryear{{Verbunt}}{{Verbunt}}{2001}]{Verbunt-2001}
{Verbunt} F.,  2001, \mn@doi [\aap] {10.1051/0004-6361:20000469}, \href
  {http://adsabs.harvard.edu/abs/2001A%26A...368..137V} {368, 137}

\bibitem[\protect\citeauthoryear{{Verner}, {Ferland}, {Korista}  \&
  {Yakovlev}}{{Verner} et~al.}{1996}]{Verner-1996}
{Verner} D.~A.,  {Ferland} G.~J.,  {Korista} K.~T.,   {Yakovlev} D.~G.,  1996,
  \mn@doi [\apj] {10.1086/177435}, \href
  {http://adsabs.harvard.edu/abs/1996ApJ...465..487V} {465, 487}

\bibitem[\protect\citeauthoryear{{Walsh}, {Cackett}  \& {Bernardini}}{{Walsh}
  et~al.}{2015}]{Walsh-2015}
{Walsh} A.~R.,  {Cackett} E.~M.,   {Bernardini} F.,  2015, \mn@doi [\mnras]
  {10.1093/mnras/stv315}, \href
  {http://adsabs.harvard.edu/abs/2015MNRAS.449.1238W} {449, 1238}

\bibitem[\protect\citeauthoryear{{Watts} et~al.,}{{Watts}
  et~al.}{2016}]{Watts-2016}
{Watts} A.~L.,  et~al., 2016, \mn@doi [Reviews of Modern Physics]
  {10.1103/RevModPhys.88.021001}, \href
  {http://adsabs.harvard.edu/abs/2016RvMP...88b1001W} {88, 021001}

\bibitem[\protect\citeauthoryear{{Webb} \& {Barret}}{{Webb} \&
  {Barret}}{2007}]{Webb-2007}
{Webb} N.~A.,  {Barret} D.,  2007, \mn@doi [\apj] {10.1086/522877}, \href
  {http://adsabs.harvard.edu/abs/2007ApJ...671..727W} {671, 727}

\bibitem[\protect\citeauthoryear{{Wilms}, {Allen}  \& {McCray}}{{Wilms}
  et~al.}{2000}]{Wilms-2000}
{Wilms} J.,  {Allen} A.,   {McCray} R.,  2000, \mn@doi [\apj] {10.1086/317016},
  \href {http://adsabs.harvard.edu/abs/2000ApJ...542..914W} {542, 914}

\bibitem[\protect\citeauthoryear{{Zampieri}, {Turolla}, {Zane}  \&
  {Treves}}{{Zampieri} et~al.}{1995}]{Zampieri-1995}
{Zampieri} L.,  {Turolla} R.,  {Zane} S.,   {Treves} A.,  1995, \mn@doi [\apj]
  {10.1086/175223}, \href {http://adsabs.harvard.edu/abs/1995ApJ...439..849Z}
  {439, 849}

\makeatother
\end{thebibliography}







\bsp	
\label{lastpage}
\end{document}